\documentclass[aps,prstab,twocolumn,showpacs,showkeys,groupedaddress]{revtex4-1}

\pdfoutput=0

\usepackage{graphicx} 
\usepackage{siunitx}
\usepackage{multirow}
\usepackage[english]{babel}
\usepackage{tablefootnote}
\usepackage{natbib}

\usepackage[dvipdfm,hidelinks,breaklinks=true,pdfborder={0 0 0},colorlinks]{hyperref}
\hypersetup{
    linkcolor={black},
    citecolor={black},
    urlcolor={black}
}

\newcommand{\sub}[1]{\ensuremath{_{\mbox{\protect\scriptsize{#1}}}}}
\newcommand{\minitab}[2][l]{\begin{tabular}{#1}#2\end{tabular}}

\usepackage{fancyhdr}
\begin{document}

\onecolumngrid
\begin{center}
\small{\textit{Submitted to PRST-AB on July 16, 2015}}
\end{center}

\title{Depinning of Trapped Magnetic Flux in Bulk Niobium SRF Cavities}

\author{S. Aull}
\email[]{sarah.aull@cern.ch}
\affiliation{CERN, Geneva, Switzerland}
\affiliation{Universit{\"a}t Siegen, Germany}

\author{J. Knobloch}
\email[]{jens.knobloch@helmholtz-berlin.de}
\affiliation{Helmholtz-Zentrum Berlin, Germany}
\affiliation{Universit{\"a}t Siegen, Germany}

\date{\today}

\begin{abstract}
Trapped magnetic flux is known to be a significant contribution to the residual resistance of superconducting radio frequency (SRF) cavities. 
The additional losses depend strongly if the vortices are depinned by the RF. 
The depinning is affected by the purity of the material, and the size of the pinning centers, as well as the cavity operation frequency. One may define a depinning frequency, above which significant depinning occurs.
This publication presents a derivation of the depinning frequency from experimental data. We find a depinning frequency of \SI{673}{MHz} for RRR \num{110} niobium. On this basis the currently used model is extended to describe the trapped flux sensitivity as function of residual resistance ratio (RRR) and operation frequency while also accounting for the pinning center size and the treatment history of the cavity. 
Moreover, the model offers an explanation for the significantly higher trapped flux sensitivity reported for nitrogen doped and \SI{120}{\degree} baked cavities.
\end{abstract}


\maketitle

\section{Introduction}
Particle accelerators operating superconducting radio frequency (SRF) cavities in continuous wave (cw) mode or at high duty cycle require maximum quality factor rather than highest accelerating gradients. 
It is therefore mandatory to minimize any residual losses and hence trapped magnetic field which is one significant contributor to the residual resistance.
While the BCS resistance decreases exponentially with temperature, the residual is temperature independent and becomes easily the dominant contribution at typical operation temperatures between \SI{1.8}{K} and \SI{2}{K}.

When an SRF cavity passes through the superconducting transition all ambient magnetic field should be expelled (Meissner effect). 
If no specific measures are taken, this expulsion is usually not complete and the magnetic flux is pinned at imperfections in the crystal lattice or impurities.
These trapped vortices have a normal conducting core and dissipate power when oscillating under the influence of the RF field.
Vallet and co-workers proposed a simple model to describe these additional losses due to trapped flux \citep{vallet_residual_1993}:

\begin{equation}\label{eq:oldscaling}
	R\sub{TF} =  \SI{3.6}{\nano\ohm/\micro\tesla}\sqrt{\frac{f}{RRR}}
\end{equation}

They consider the normal electrons in the core of a flux tube being lossy due to the skin effect which scales with the square root of the frequency $f$ and with the square root of the electrical conductivity which in turn is inversely proportional to the residual resistance ratio $RRR$.
The sensitivity of \SI{3.6}{\nano\ohm/\micro\tesla} was derived experimentally using a \SI{1.5}{GHz} cavity made from RRR \num{300} material.
This simple model holds however only if the trapped vortices actually move under the influence of the RF field.

\begin{figure}
	\centering
		\includegraphics[width=0.8\linewidth]{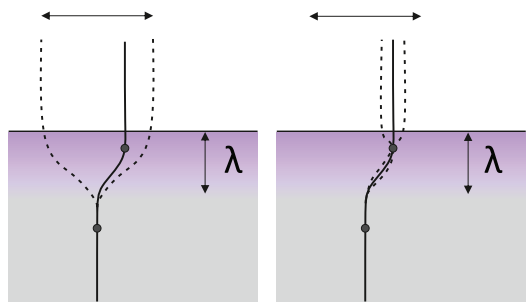}
	\caption{Flux line motion under the influence of an RF field. The solid line represents a single flux line trapped by two pinning centres. The dashed lines indicate the range of motion for the flux line assuming a bulk superconductor with $f/f\sub{0} \gg 1$ (left) and a bulk superconductor with $f/f\sub{0} \ll 1$ (right).}\label{fig:depinning}
\end{figure}

The displacement of a single flux line can be described as a damped oscillator accounting for the Lorentz force due to a transport current $J\sub{T}$, the frictional force and the pinning force \citep{gittleman_radio-frequency_1966,shapira_magnetoacoustic_1967,gittleman_pinning_1968}:

\begin{equation}
	m\ddot{x} + \eta\dot{x}+kx = \frac{J\sub{T}\Phi\sub{0}}{c}
\end{equation}

Here $m$ is the effective mass of the vortex.
The contribution from this term can be neglected as the RF period ($>\SI{E-8}{s}$) is much larger than the relaxation time of a vortex $\tau\sim m$, which was estimated by H. Suhl to be in the order of \SI{E-12}{s} \citep{shapira_magnetoacoustic_1967,suhl_inertial_1965}.
$k$ accounts for the pinning force and depends on the size, geometry and type of the pinning centre.

The flux flow viscosity $\eta$ is given by the flux quantum $\Phi\sub{0}$ divided by the squared speed of light $c^2$, the upper critical field $H\sub{c2}$ and the electrical resistivity $\rho\sub{n}$ \citep{strnad_dissipative_1964}:

\begin{equation}\label{eq:eta}
	\eta = \frac{\Phi\sub{0}H\sub{c2}}{c^2 \rho\sub{n}}.
\end{equation}

The depinning frequency $f\sub{0}$ is defined as the frequency where \SI{50}{\%} of the pinned vortices are depinned and is given by the ratio of pinning constant and flux flow viscosity \citep{schmid_theory_1973}:

\begin{equation}
	f\sub{0} = \frac{k}{\eta}
\end{equation}

\begin{figure}
	\centering
		\includegraphics[width=0.94\linewidth]{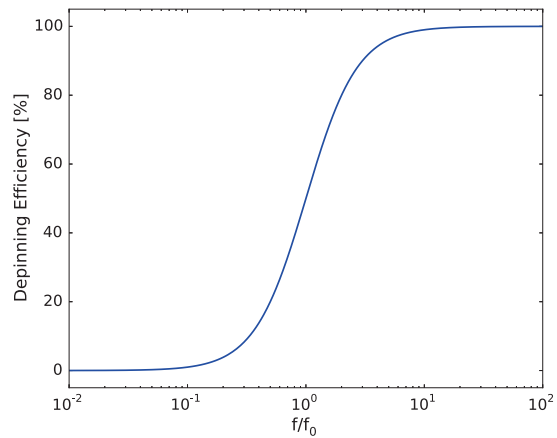}
	\caption{Depinning efficiency as function of the ratio between operating and depinning frequency. For $f/f\sub{0} \gg 1$ all flux lines are depinned while for $f/f\sub{0} \ll 1$ all flux lines are pinned and do not dissipate power.}\label{fig:depinning_efficiency}
\end{figure}

If the RF frequency is well above the depinning frequency, the majority of trapped vortices are depinned. However, if the superconductor is thicker than the penetration depth $\lambda$, as it is the case for SRF cavities, the trapped vortices are only depinned in the RF layer and stay pinned in the bulk. They are then oscillating within the RF layer but cannot move entirely freely or be expelled from the material. This scenario is depicted in Fig.~\ref{fig:depinning} on the left.
If the RF frequency is well below the depinning frequency, the pinning is effective and the flux line can only move in between pinning centers as indicated in Fig.~\ref{fig:depinning} on the right.
Based on Gittleman and Rosenblum's calculation of the impedance of samples thinner than the penetration depth \citep{gittleman_pinning_1968}, we define a depinning efficiency for bulk superconductors to describe how many flux lines are depinned depending on the ratio of RF frequency to depinning frequency  and plot it in Figure \ref{fig:depinning_efficiency}:

\begin{equation}\label{eq:depin_efficiency}
	\varepsilon\sub{depin} = \frac{f^2}{f^2+f\sub{0}^2} 
\end{equation}

Since $\rho\sub{n}$ is inversely proportional to the conductivity and therefore to the RRR, the depinning frequency scales with $1/RRR$. 
Considering typical SRF cavities, we can then identify two distinct regimes in Fig.~\ref{fig:depinning_efficiency}: 
A lower regime where $f/f\sub{0}\ll 1$ so that pinning is very efficient and the upper part where $f/f\sub{0} \gg 1$ so that all flux lines are depinned. The first can be obtained by the combination of low RF frequency and low RRR (= large $f\sub{0}$) as typical for niobium coated cavities. 
The latter represents the combination of high RF frequency and high RRR (=low $f\sub{0}$) as typical for bulk niobium cavities.
This could already explain why niobium coated cavities are at least one to two orders of magnitude less sensitive to trapped flux as bulk \citep{benvenuti_study_1999,zhang_influence_2014} so that they do not require magnetic shielding \citep{bruning_lhc_2004}.

\section{Depinning Frequency}

Bulk niobium cavities are nowadays made from high RRR ($>\num{250}$) niobium and are designed for rather high operation frequencies ($>\SI{1}{GHz}$). 
This combination usually places the cavities on the upper plateau of the depinning curve in Fig.~\ref{fig:depinning_efficiency}. 
Arnolds-Mayer and Chiaveri published in 1987 trapped flux measurements on a \SI{500}{MHz} cavity of RRR \num{110} material \citep{arnolds-mayer_500_1987}.
The authors derive a trapped flux sensitivity of \SI{1.22}{\nano\ohm/\micro\tesla}. 
Following Eq. \ref{eq:oldscaling} as proposed by Vallet and co-workers, a trapped flux sensitivity of \SI{3.4}{\nano\ohm/\micro\tesla} would be expected. 
Comparing the measurement of Arnolds-Mayer and Chiaveri with the prediction of Vallet et. al. yields a depinning efficiency of \SI{35.5}{\%} placing this measurement on the slope between the pinning and depinning regime.
Equation \ref{eq:depin_efficiency} can now be used to calculate the depinning frequency for RRR \num{110}:

\begin{equation}\label{eq:f0}
f\sub{0} = \sqrt{\frac{f^2\left( \varepsilon\sub{pin}-1\right)}{\varepsilon\sub{pin}}} = \SI{673}{MHz}
\end{equation}

\begin{figure}[htb]
	\centering
		\includegraphics[width=0.94\linewidth]{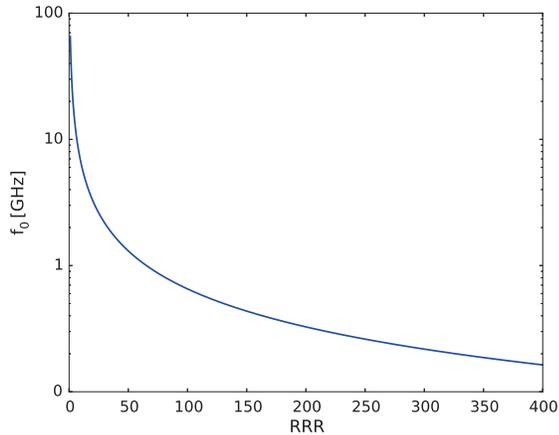}
	\caption{Depinning frequency as function of RRR.}
	\label{fig:f0vsRRR}
\end{figure}

For other RRR values the depinning frequency can be scaled via $1/RRR$. Figure \ref{fig:f0vsRRR} displays the change of $f\sub{0}$ with RRR. 
As can be seen, the depinning frequency increases dramatically for low RRR values. For niobium films a RRR of \num{10} to \num{20} is typical. 
The depinning frequency for this range is between \SI{3.7}{GHz} and \SI{7.4}{GHz}.
This is consistent with niobium film cavities being almost insensitive to trapped flux even at \SI{1.5}{GHz} \citep{benvenuti_study_1999}. 
In contrast to the niobium films, the depinning frequency for RRR \num{300} is \SI{247}{MHz}.

\section{Trapped Flux Sensitivity}

In order to describe additional residual resistance due to trapped flux, we account for three aspects:
The losses of depinned vortices under the influence of the RF field, the percentage of vortices that get depinned and the total amount of trapped field in the material.
The first factor is well described by Vallet and co-workers with dependency on frequency and RRR according to the normal skin effect. 
The second factor is given be the depinning efficiency as described in Eq. \ref{eq:depin_efficiency} and the depinning frequency can be scaled from Eq. \ref{eq:f0}. 
However, not only the RRR defines the depinning frequency but also the size of the pinning center. Larkin and Ovchinnikov \citep{larkin_pinning_1979} discuss pinning of a single vortex parallel to a grain boundary. They conclude that the depinning frequency increases linearly with the thickness of the grain boundary. Flux trapping measurements on bulk niobium samples suggest that grain boundaries are the most severe pinning centers in cavity material \citep{aull_trapped_2012}.
Based on Larkin and Ovchinnikovs calculations, we introduce a relative pinning center size $p$ which is normalized to standard high quality bulk niobium as nowadays used for SRF cavities.

Lastly, the amount of trapped flux has to be estimated. It has been shown that the residual resistance increases linearly with the applied (dc) magnetic field $B\sub{app}$ up to \num{6} times the earth magnetic field strength \citep{vallet_flux_1992,aull_trapped_2012}.
However, the treatment history and the crystal structure of the material has an influence on how much of the ambient field actually gets trapped \citep{aull_trapped_2012}.
Moreover, recent studies show that acting on the cool down dynamics at the superconducting transition can lead to improved flux expulsion \citep{aull_trapped_2012,vogt_impact_2013,romanenko_dependence_2014}.
In the following, we will restrict the discussions to trapped flux measurements where the additional residual resistance due to trapped flux is high enough so that cool down effects can be neglected. 
Moreover, we restrict ourselves to cavity measurement cooled down in an axial field since it has been recently shown that the direction of the field has also an influence on the trapped flux sensitivity \citep{checchin_origin_2015}.  

\begin{widetext}
\begin{equation}\label{eq:Rs(f,RRR)}
S\sub{TF} = \SI{3.6}{\nano\ohm/\micro\tesla}\sqrt{\frac{f}{\SI{1.5}{GHz}}\frac{300}{RRR}}\cdot\frac{f^2}{f^2+\left(p\cdot\SI{673}{MHz}\frac{110}{RRR}\right)^2}\cdot \varepsilon\sub{trap}
\end{equation}
\end{widetext}

The effect of the treatment history and the base material (large grain vs fine grain) are described by a trapping efficiency $\varepsilon\sub{trap}$ and can be taken from \citep{aull_trapped_2012} for several material and treatment combinations.
Folding these three aspects into a trapped flux sensitivity results in Eq. \ref{eq:Rs(f,RRR)} which can be tested against SRF measurements. The corresponding total increase of residual resistance due to trapped flux is then $R\sub{TF} = S\sub{TF}\cdot B\sub{app}$.

Table \ref{tab:TFsensitivity} lists available trapped flux measurements where the effect of cooling can be neglected and sufficient information about the treatment and the cavity material is reported. 
Depending on the treatment, a flux trapping efficiency $\varepsilon\sub{trap}$ is assumed based on \citep{aull_trapped_2012} and listed in Tab. \ref{tab:TFsensitivity}.
It can be seen that Eq. \ref{eq:Rs(f,RRR)} agrees well with the available cavity data for medium and high RRR values.
The effect of RRR, frequency and relative pinning center size on the trapped flux sensitivity will be subject in the following sections.

\begin{table*}
\caption{Measured trapped flux sensitivity for different bulk Nb cavities.
The measurements are compared with the calculation according to Eq. \ref{eq:Rs(f,RRR)}. Frequency and RRR are taken from the references. If the mean free path $\ell$ is given, the RRR was calculated as  $ \ell [\si{nm}] = 2.7\cdot RRR$. The relative pinning center size $p$ was set to \num{1} for all listed calculations. See text for discussion on $p$ for the N doped and \SI{120}{\degree} baked cavities.}
\begin{ruledtabular}
	\begin{tabular}{ccccccc}
	\multirow{2}{*}{}	&\multirow{2}{*}{RRR}		& \multirow{2}*{\minitab[c]{Frequency \\ $[\si{GHz}]$ }}& meas. trapped flux 	& calc. trapped flux 	& \multirow{2}{*}{$\varepsilon\sub{trap}$}	& \multirow{2}{*}{Ref.}\\
				&				&			&sensitivity [\si{\nano\ohm/\micro\tesla}]	& sensitivity	[\si{\nano\ohm/\micro\tesla}]&&\\\hline
	Cavity		&	\num{300}	& 1.5		& 3.6	&	3.5	&	\SI{100}{\%} & \citep{vallet_residual_1993}\\
	Cavity		&	\num{110}	& 0.5		& 1.22	&	1.22	& \SI{100}{\%} & \citep{arnolds-mayer_500_1987} \\
	Large grain cavity 	+ BCP &	200	&1.5		& $2.5 \pm 0.2	$& 3.0 &	\SI{73}{\%}	& \citep{ciovati_measurement_2007}  \\
	Large grain cavity 	&	\multirow{2}{*}{\num{200}}	&\multirow{2}{*}{\num{1.5}}		& \multirow{2}{*}{$1.43 \pm 0.12$}& \multirow{2}{*}{1.75} &	\multirow{2}{*}{\SI{42}{\%}}	&  \multirow{2}{*}{\citep{ciovati_measurement_2007}} \\
	+ \SI{1250}{\celsius}  + BCP	&\\
	N doped cavity & \num{3.3} & \num{1.3} & \num{11.3}	& \num{0.1}\footnote{The calculation can be brought to an agreement with the measurement for $p=0.08$. See text for discussion.}	& \SI{100}{\%} & \citep{gonella_flux_2014}\\
	\SI{120}{\degree} baked cavity & \num{8.5} & \num{1.3}& \num{3.7}	& \num{0.4}\footnote{The calculation can be brought to an agreement with the measurement for $p=0.31$. See text for discussion.}	& \SI{100}{\%} & \citep{gonella_flux_2014}
	\end{tabular}
	\label{tab:TFsensitivity}
	\end{ruledtabular}
\end{table*}

\subsection{The Influence of RRR}

\begin{figure}
\centering
\includegraphics[width=0.94\linewidth]{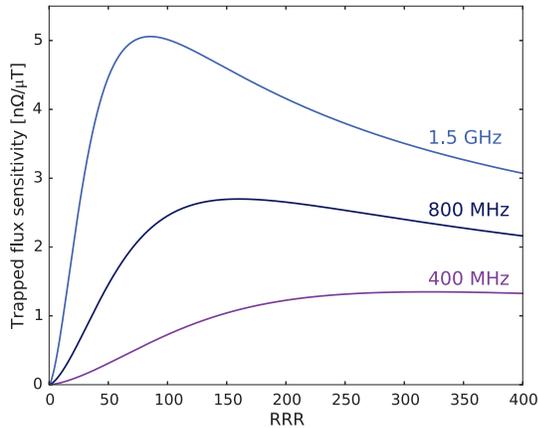}
\caption{Trapped flux sensitivity as function of RRR for different frequencies.}\label{fig:TFsens_RRR}
\end{figure}

Figure \ref{fig:TFsens_RRR} shows the trapped flux sensitivity for different cavity frequencies as function of RRR. 
For high RRR values and low frequency the trapped flux sensitivity is basically constant. 
For high frequencies however, there is slight decrease with increasing RRR. 
The measurements on large grain niobium reported in \citep{ciovati_measurement_2007} list a RRR of \num{200} before baking.
It can be expected that the baking at \SI{1250}{\celsius} increased the RRR due to the post-purification process \citep{ciovati_measurement_2007}. 
The prediction can be brought to agreement with the measurement for a RRR value of \num{320} which appears plausible.
Moreover, Fig.~\ref{fig:TFsens_RRR} reveals an increase of trapped flux sensitivity when going from high RRR to intermediate values and a rapid decrease for very low RRR values.
The very strong change in trapped flux sensitivity for low RRR values, especially at high frequency requires therefore information about the surface RRR to make a reliable prediction.
  
It should also be noted that the scaling with $\sqrt{f/RRR}$ for the losses as described by Eq. \ref{eq:oldscaling} are based on the normal skin effect. 
For high RRR values the mean free path becomes comparable to the skin depth and the anomalous skin effect has to be applied. In this regime, the losses in the normal conducting core become independent of the RRR and scale with $f^{2/3}$.
 
\subsection{The Influence of the Pinning Center Size}\label{sec:p}

\begin{figure}
\centering
\includegraphics[width=0.94\linewidth]{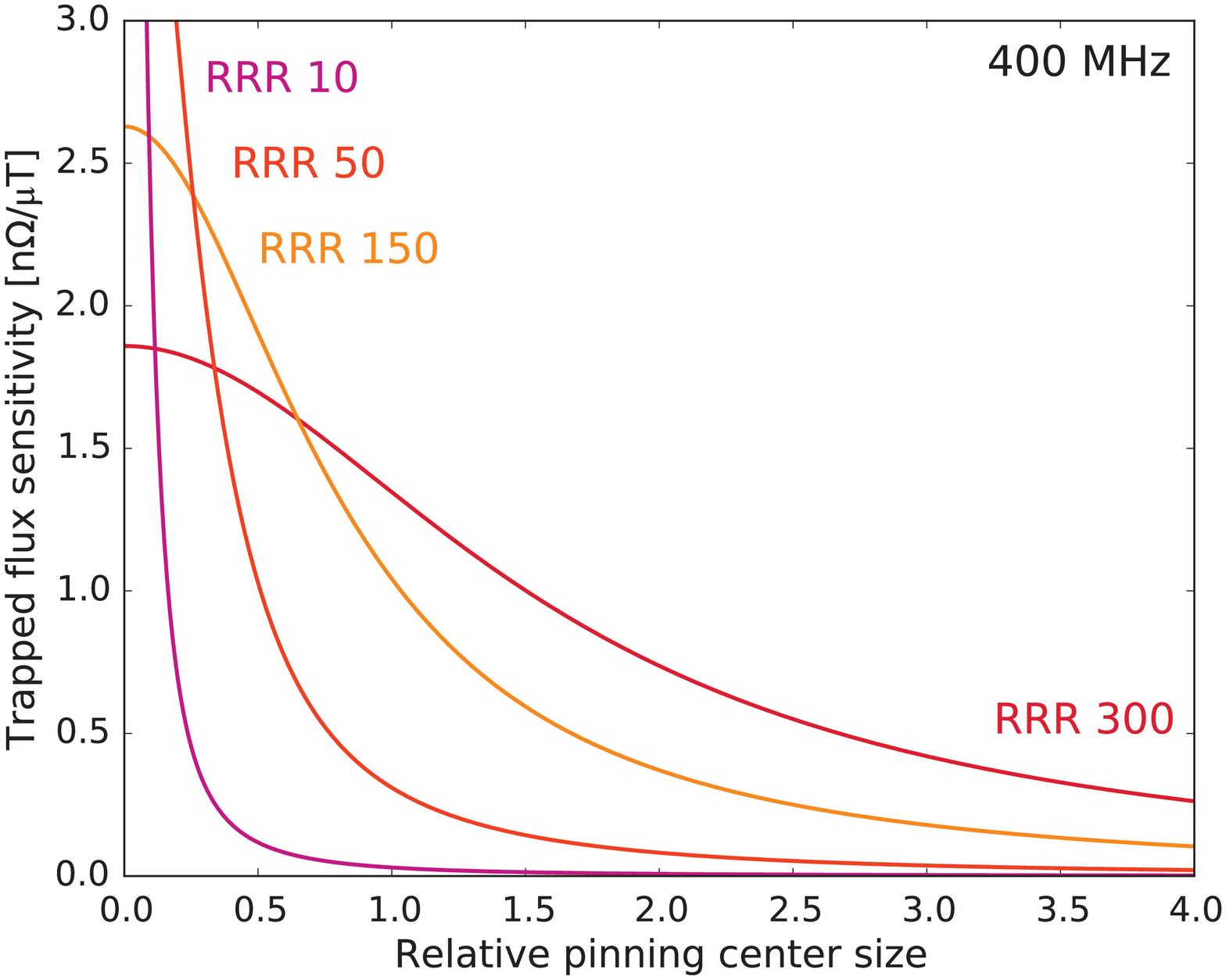}
\caption{Trapped flux sensitivity as function of pinning center size for \SI{400}{MHz}.}\label{fig:TFsens_p_400}
\end{figure}

\begin{figure}
\centering
\includegraphics[width=0.94\linewidth]{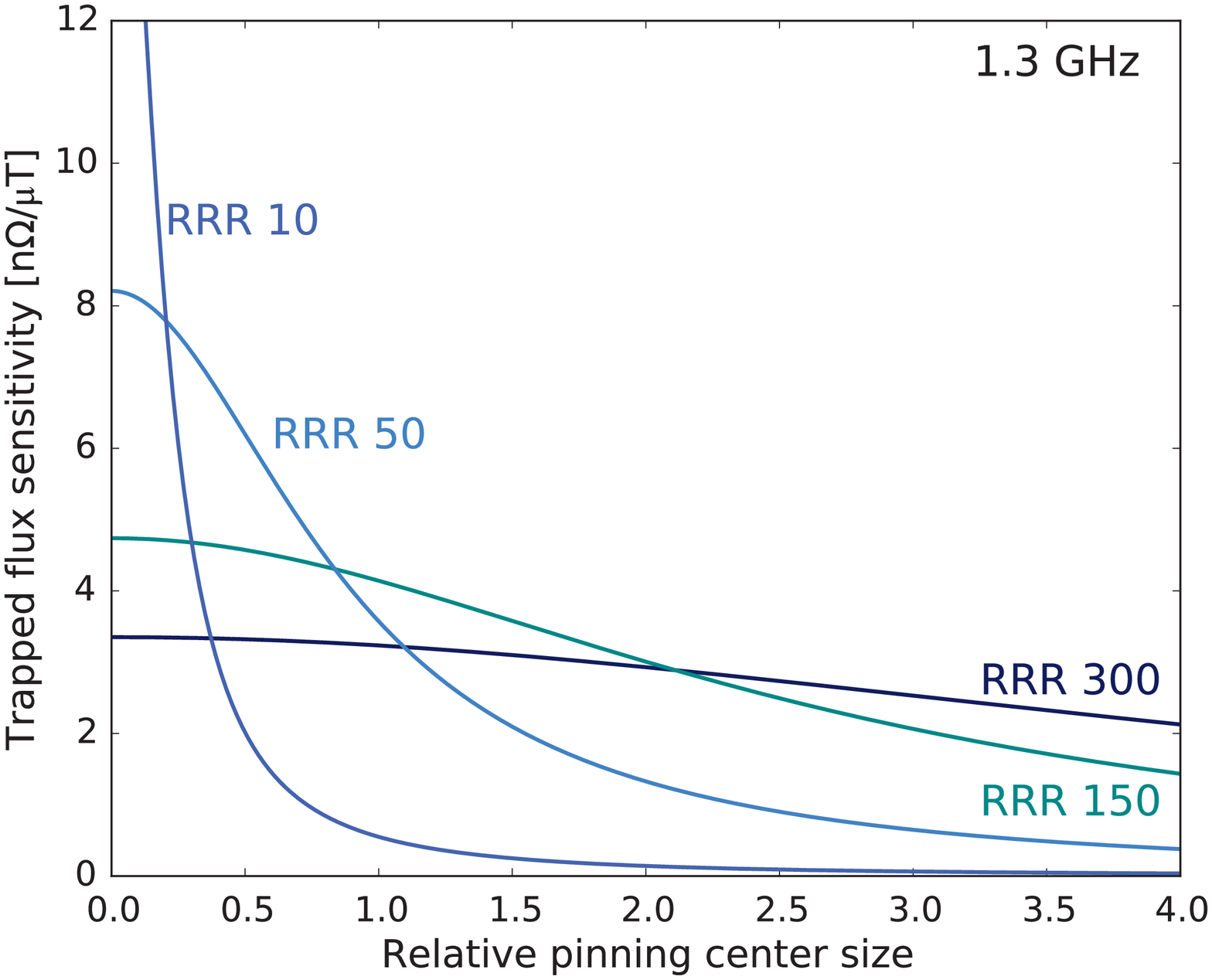}
\caption{Trapped flux sensitivity as function of pinning center size for \SI{1.3}{GHz}.}\label{fig:TFsens_p_1300}
\end{figure}

Figures \ref{fig:TFsens_p_400} and \ref{fig:TFsens_p_1300} plot the trapped flux sensitivity as function of the relative pinning center size $p$ for \SI{400}{MHz} and \SI{1.3}{GHz}.
The trend for both frequencies is very similar: for high RRR the trapped flux sensitivity depends only slightly on the pinning center size. 
The dependency however becomes stronger the shorter the mean free path gets and the effect is more emphasized for high frequencies.

There are several trapped flux measurements at \SI{1.3}{GHz} with a base material of RRR \num{300} which where baked at \SI{120}{\degree} or nitrogen doped \citep{gonella_flux_2014}. 
Both treatments result in a drastic reduction of surface RRR.
The measured trapped flux sensitivity of both cavities is significantly higher compared to what would be expected from either Eq. \ref{eq:oldscaling} or Eq. \ref{eq:Rs(f,RRR)}.
Recent findings suggest that the \SI{120}{\degree} baking as well as the nitrogen doping reduces or even prohibits the formation of nanohydrides which most likely act as pinning centers \citep{yulia_trenikhina_tem_2014,trenikhina_nanostructural_2015}.
Moreover, it is suggested that the size of the nanohydrides is reduced.
Assuming nanohydrides to be pinning centers, our model supports a reduced pinning center size: 
An agreement with the measurements can be obtained for a relative pinning center size of \num{0.08} for the N doping and \num{0.31} for the \SI{120}{\degree} baking.

\begin{figure}
\centering
\includegraphics[width=0.94\linewidth]{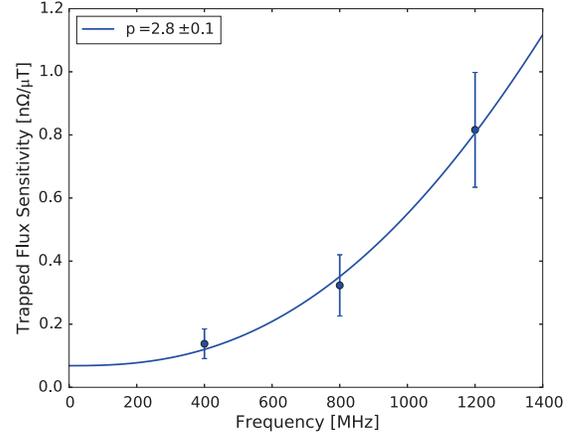}
\caption{Trapped flux sensitivity measured with the Quadrupole Resonator for \SIlist{400;800;1200}{MHz} of a RRR \num{47} bulk Nb sample.}\label{fig:TF_sens_QPR}
\end{figure}

The relative pinning center size can also be estimated from measurements with the Quadrupole Resonator which allows measurements of the surface resistance for \SIlist{400;800;1200}{MHz} with the same field configuration. The full description of the set-up, the measurement method and parameter range can be found elsewhere \citep{junginger_extension_2012}. 
Recent upgrades allow applying a dc magnetic field to the sample under test. The field strength is measured with a cryogenic magnetic field probe (Mag-01H \citep{mag-01h_2014}).
In a dedicated test, a reactor grade bulk Nb sample with RRR \num{47} was cooled down in different ambient fields and the surface resistance was measured at \SI{2.5}{K} and low RF field ($\approx \SI{10}{mT}$). For each trapped field, the surface resistance was measured at each of the three resonant frequencies (\SIlist{400;800;1200}{MHz}). 
In this way, the trapped flux sensitivity was derived as function of frequency and is plotted in Fig.~\ref{fig:TF_sens_QPR}.
The relative pinning center size can now be fitted using Eq. \ref{eq:Rs(f,RRR)}. 
We find excellent agreement for $p=(2.8 \pm 0.1)$.
The relatively bigger pinning center size for this sample appears plausible as the base material is reactor grade niobium.
This is in contrast to measurements on cavities where the base material was of high quality and the mean free path was shortened due to diffusion processes of dissolved gases upon baking.
Our measurement supports therefore the validity of Eq. \ref{eq:Rs(f,RRR)} and the assumptions made.

\section{Summary and Conclusion}
Depinning of trapped vortices has to be considered when predictions for the trapped flux sensitivity of SRF cavities should be made.
The depinning efficiency depends on the size of the pinning centers and the RF frequency.
Trapped flux measurements on a low frequency and medium RRR cavity allowed deriving the according depinning frequency which can be scaled to other RRR values.
On this basis we propose an extended model for describing the additional losses due to trapped flux.
To date, only the losses in the normal core due to the skin effect have been taken into account. 
We account additionally for depinning which depends on the RRR, operation frequency and the pinning center size as well as for the trapping efficiency.
The predictions made are in good agreement with our measurements and measurements done at other labs covering a variety of frequency, RRR and treatment combinations.
Moreover, the model offers an explanation why cavities with short mean free path are much more sensitive to trapped flux than the simple model suggests.

\begin{acknowledgments}
We would like to acknowledge V. Palmieri for calling our attention to the publications on depinning. We thank W. Venturini Delsolaro and T. Junginger for fruitful discussions and help with the Quadrupole Resonator.
Moreover, we thank J. Bremer, L. Dufay-Chanat and T. Koettig for providing the cryogenic infrastructure and technical help, as well as G.
Pechaud and S. Forel for preparing the sample for the tests in the Quadrupole Resonator.
This work is sponsored by the Wolfgang Gentner Programme of the German Federal Ministry of Education and Research (BMBF).
\end{acknowledgments}

\bibliographystyle{unsrt}

\bibliography{References}

\end{document}